\documentclass[aps,prl,reprint,superscriptaddress,
amsfonts,amssymb,amsmath]{revtex4-1}

\usepackage{dcolumn}
\usepackage{graphicx}
\usepackage[colorlinks,
            linkcolor=red,
            anchorcolor=blue,
            citecolor=green,
            ]{hyperref}

\newcommand{\mlb}{$M_{\ell j}$}
\newcommand{\mlvb}{$M_{\ell \nu j}$}

\newcommand{\pt}{$p_{T}$}
\newcommand{\et}{$E_{T}$}

\newcommand{\gevcc}{GeV/$c^2$}
\newcommand{\gevc}{GeV/$c$}
\newcommand{\MET}{\mbox{$\raisebox{.3ex}{$\not\!$}E_T$}}
\newcommand{\METVEC}{\mbox{$\raisebox{.3ex}{$\not\!$}{\vec E}_T$}}

\begin{document}


\title{Evidence for $s$-channel Single-Top-Quark Production in Events with one Charged
  Lepton and two Jets at CDF}



\affiliation{Institute of Physics, Academia Sinica, Taipei, Taiwan 11529, Republic of China}
\affiliation{Argonne National Laboratory, Argonne, Illinois 60439, USA}
\affiliation{University of Athens, 157 71 Athens, Greece}
\affiliation{Institut de Fisica d'Altes Energies, ICREA, Universitat Autonoma de Barcelona, E-08193, Bellaterra (Barcelona), Spain}
\affiliation{Baylor University, Waco, Texas 76798, USA}
\affiliation{Istituto Nazionale di Fisica Nucleare Bologna, \ensuremath{^{ii}}University of Bologna, I-40127 Bologna, Italy}
\affiliation{University of California, Davis, Davis, California 95616, USA}
\affiliation{University of California, Los Angeles, Los Angeles, California 90024, USA}
\affiliation{Instituto de Fisica de Cantabria, CSIC-University of Cantabria, 39005 Santander, Spain}
\affiliation{Carnegie Mellon University, Pittsburgh, Pennsylvania 15213, USA}
\affiliation{Enrico Fermi Institute, University of Chicago, Chicago, Illinois 60637, USA}
\affiliation{Comenius University, 842 48 Bratislava, Slovakia; Institute of Experimental Physics, 040 01 Kosice, Slovakia}
\affiliation{Joint Institute for Nuclear Research, RU-141980 Dubna, Russia}
\affiliation{Duke University, Durham, North Carolina 27708, USA}
\affiliation{Fermi National Accelerator Laboratory, Batavia, Illinois 60510, USA}
\affiliation{University of Florida, Gainesville, Florida 32611, USA}
\affiliation{Laboratori Nazionali di Frascati, Istituto Nazionale di Fisica Nucleare, I-00044 Frascati, Italy}
\affiliation{University of Geneva, CH-1211 Geneva 4, Switzerland}
\affiliation{Glasgow University, Glasgow G12 8QQ, United Kingdom}
\affiliation{Harvard University, Cambridge, Massachusetts 02138, USA}
\affiliation{Division of High Energy Physics, Department of Physics, University of Helsinki, FIN-00014, Helsinki, Finland; Helsinki Institute of Physics, FIN-00014, Helsinki, Finland}
\affiliation{University of Illinois, Urbana, Illinois 61801, USA}
\affiliation{The Johns Hopkins University, Baltimore, Maryland 21218, USA}
\affiliation{Institut f\"{u}r Experimentelle Kernphysik, Karlsruhe Institute of Technology, D-76131 Karlsruhe, Germany}
\affiliation{Center for High Energy Physics: Kyungpook National University, Daegu 702-701, Korea; Seoul National University, Seoul 151-742, Korea; Sungkyunkwan University, Suwon 440-746, Korea; Korea Institute of Science and Technology Information, Daejeon 305-806, Korea; Chonnam National University, Gwangju 500-757, Korea; Chonbuk National University, Jeonju 561-756, Korea; Ewha Womans University, Seoul, 120-750, Korea}
\affiliation{Ernest Orlando Lawrence Berkeley National Laboratory, Berkeley, California 94720, USA}
\affiliation{University of Liverpool, Liverpool L69 7ZE, United Kingdom}
\affiliation{University College London, London WC1E 6BT, United Kingdom}
\affiliation{Centro de Investigaciones Energeticas Medioambientales y Tecnologicas, E-28040 Madrid, Spain}
\affiliation{Massachusetts Institute of Technology, Cambridge, Massachusetts 02139, USA}
\affiliation{University of Michigan, Ann Arbor, Michigan 48109, USA}
\affiliation{Michigan State University, East Lansing, Michigan 48824, USA}
\affiliation{Institution for Theoretical and Experimental Physics, ITEP, Moscow 117259, Russia}
\affiliation{University of New Mexico, Albuquerque, New Mexico 87131, USA}
\affiliation{The Ohio State University, Columbus, Ohio 43210, USA}
\affiliation{Okayama University, Okayama 700-8530, Japan}
\affiliation{Osaka City University, Osaka 558-8585, Japan}
\affiliation{University of Oxford, Oxford OX1 3RH, United Kingdom}
\affiliation{Istituto Nazionale di Fisica Nucleare, Sezione di Padova, \ensuremath{^{jj}}University of Padova, I-35131 Padova, Italy}
\affiliation{University of Pennsylvania, Philadelphia, Pennsylvania 19104, USA}
\affiliation{Istituto Nazionale di Fisica Nucleare Pisa, \ensuremath{^{kk}}University of Pisa, \ensuremath{^{ll}}University of Siena, \ensuremath{^{mm}}Scuola Normale Superiore, I-56127 Pisa, Italy, \ensuremath{^{nn}}INFN Pavia, I-27100 Pavia, Italy, \ensuremath{^{oo}}University of Pavia, I-27100 Pavia, Italy}
\affiliation{University of Pittsburgh, Pittsburgh, Pennsylvania 15260, USA}
\affiliation{Purdue University, West Lafayette, Indiana 47907, USA}
\affiliation{University of Rochester, Rochester, New York 14627, USA}
\affiliation{The Rockefeller University, New York, New York 10065, USA}
\affiliation{Istituto Nazionale di Fisica Nucleare, Sezione di Roma 1, \ensuremath{^{pp}}Sapienza Universit\`{a} di Roma, I-00185 Roma, Italy}
\affiliation{Mitchell Institute for Fundamental Physics and Astronomy, Texas A\&M University, College Station, Texas 77843, USA}
\affiliation{Istituto Nazionale di Fisica Nucleare Trieste, \ensuremath{^{qq}}Gruppo Collegato di Udine, \ensuremath{^{rr}}University of Udine, I-33100 Udine, Italy, \ensuremath{^{ss}}University of Trieste, I-34127 Trieste, Italy}
\affiliation{University of Tsukuba, Tsukuba, Ibaraki 305, Japan}
\affiliation{Tufts University, Medford, Massachusetts 02155, USA}
\affiliation{University of Virginia, Charlottesville, Virginia 22906, USA}
\affiliation{Waseda University, Tokyo 169, Japan}
\affiliation{Wayne State University, Detroit, Michigan 48201, USA}
\affiliation{University of Wisconsin, Madison, Wisconsin 53706, USA}
\affiliation{Yale University, New Haven, Connecticut 06520, USA}

\author{T.~Aaltonen}
\affiliation{Division of High Energy Physics, Department of Physics, University of Helsinki, FIN-00014, Helsinki, Finland; Helsinki Institute of Physics, FIN-00014, Helsinki, Finland}
\author{S.~Amerio\ensuremath{^{jj}}}
\affiliation{Istituto Nazionale di Fisica Nucleare, Sezione di Padova, \ensuremath{^{jj}}University of Padova, I-35131 Padova, Italy}
\author{D.~Amidei}
\affiliation{University of Michigan, Ann Arbor, Michigan 48109, USA}
\author{A.~Anastassov\ensuremath{^{v}}}
\affiliation{Fermi National Accelerator Laboratory, Batavia, Illinois 60510, USA}
\author{A.~Annovi}
\affiliation{Laboratori Nazionali di Frascati, Istituto Nazionale di Fisica Nucleare, I-00044 Frascati, Italy}
\author{J.~Antos}
\affiliation{Comenius University, 842 48 Bratislava, Slovakia; Institute of Experimental Physics, 040 01 Kosice, Slovakia}
\author{G.~Apollinari}
\affiliation{Fermi National Accelerator Laboratory, Batavia, Illinois 60510, USA}
\author{J.A.~Appel}
\affiliation{Fermi National Accelerator Laboratory, Batavia, Illinois 60510, USA}
\author{T.~Arisawa}
\affiliation{Waseda University, Tokyo 169, Japan}
\author{A.~Artikov}
\affiliation{Joint Institute for Nuclear Research, RU-141980 Dubna, Russia}
% ...

\collaboration{CDF Collaboration}
\altaffiliation[With visitors from]{
\ensuremath{^{a}}University of British Columbia, Vancouver, BC V6T 1Z1, Canada,
\ensuremath{^{b}}Istituto Nazionale di Fisica Nucleare, Sezione di Cagliari, 09042 Monserrato (Cagliari), Italy,
\ensuremath{^{c}}University of California Irvine, Irvine, CA 92697, USA,
\ensuremath{^{d}}Institute of Physics, Academy of Sciences of the Czech Republic, 182~21, Czech Republic,
\ensuremath{^{e}}CERN, CH-1211 Geneva, Switzerland,
\ensuremath{^{f}}Cornell University, Ithaca, NY 14853, USA,
\ensuremath{^{g}}University of Cyprus, Nicosia CY-1678, Cyprus,
\ensuremath{^{h}}Office of Science, U.S. Department of Energy, Washington, DC 20585, USA,
\ensuremath{^{i}}University College Dublin, Dublin 4, Ireland,
\ensuremath{^{j}}ETH, 8092 Z\"{u}rich, Switzerland,
\ensuremath{^{k}}University of Fukui, Fukui City, Fukui Prefecture, Japan 910-0017,
\ensuremath{^{l}}Universidad Iberoamericana, Lomas de Santa Fe, M\'{e}xico, C.P. 01219, Distrito Federal,
\ensuremath{^{m}}University of Iowa, Iowa City, IA 52242, USA,
\ensuremath{^{n}}Kinki University, Higashi-Osaka City, Japan 577-8502,
\ensuremath{^{o}}Kansas State University, Manhattan, KS 66506, USA,
\ensuremath{^{p}}Brookhaven National Laboratory, Upton, NY 11973, USA,
\ensuremath{^{q}}Queen Mary, University of London, London, E1 4NS, United Kingdom,
\ensuremath{^{r}}University of Melbourne, Victoria 3010, Australia,
\ensuremath{^{s}}Muons, Inc., Batavia, IL 60510, USA,
\ensuremath{^{t}}Nagasaki Institute of Applied Science, Nagasaki 851-0193, Japan,
\ensuremath{^{u}}National Research Nuclear University, Moscow 115409, Russia,
\ensuremath{^{v}}Northwestern University, Evanston, IL 60208, USA,
\ensuremath{^{w}}University of Notre Dame, Notre Dame, IN 46556, USA,
\ensuremath{^{x}}Universidad de Oviedo, E-33007 Oviedo, Spain,
\ensuremath{^{y}}CNRS-IN2P3, Paris, F-75205 France,
\ensuremath{^{z}}Universidad Tecnica Federico Santa Maria, 110v Valparaiso, Chile,
\ensuremath{^{aa}}The University of Jordan, Amman 11942, Jordan,
\ensuremath{^{bb}}Universite catholique de Louvain, 1348 Louvain-La-Neuve, Belgium,
\ensuremath{^{cc}}University of Z\"{u}rich, 8006 Z\"{u}rich, Switzerland,
\ensuremath{^{dd}}Massachusetts General Hospital, Boston, MA 02114 USA,
\ensuremath{^{ee}}Harvard Medical School, Boston, MA 02114 USA,
\ensuremath{^{ff}}Hampton University, Hampton, VA 23668, USA,
\ensuremath{^{gg}}Los Alamos National Laboratory, Los Alamos, NM 87544, USA,
\ensuremath{^{hh}}Universit\`{a} degli Studi di Napoli Federico I, I-80138 Napoli, Italy
}
\noaffiliation




\date{\today}

\begin{abstract} 

We report evidence for \textit{s}-channel single-top-quark production in proton-antiproton
collisions at center-of-mass energy $\sqrt{s}= 1.96~\mathrm{TeV}$ using a data set that
corresponds to an integrated luminosity of $9.4~\mathrm{fb}^{-1}$ collected by the
Collider Detector at Fermilab. We select events consistent with the $s$-channel
process including two jets and one leptonically decaying $W$ boson. The observed
significance is $3.8$ standard deviations with respect to the background-only
prediction. Assuming a top-quark mass of $172.5~\mathrm{GeV}/c^2$, we measure the
$s$-channel cross section to be $1.41^{+0.44}_{-0.42}~\mathrm{pb}$.

\end{abstract}


\maketitle

In proton-antiproton collisions, top quarks can be singly produced through electroweak
interactions. This process provides a unique opportunity to test the standard model (SM)
and search for non-SM physics. Each channel of the single-top-quark process is
sensitive to different classes of SM extensions: the $s$-channel process, in which an
intermediate $W$ boson decays into a top (antitop) quark and an antibottom (bottom) quark,
is sensitive to contributions from additional heavy bosons~\cite{PhysRevD.63.014018}; the
$t$-channel process, in which a bottom quark transforms into a top quark by exchanging a
$W$ boson with another quark, is more sensitive to flavor-changing neutral
currents~\cite{PhysRevD.63.014018}. Independently studying the production rate of these
channels provides more restrictive constraints on SM extensions than just studying the
combined production rate~\footnote{The third process, \textit{tW} production, contributes
negligibly in $p\bar{p}$ collisions, given the cross section of
$0.25$~pb~\cite{PhysRevD.81.054028} .}.

Single-top-quark production was first observed independently by the CDF and D0 experiments
in 2009~\cite{PhysRevLett.103.092002, PhysRevLett.103.092001}. The $t$-channel production
was first observed in 2011 by the D0 experiment~\cite{Abazov2011313}, and confirmed in
2012 by the ATLAS~\cite{Aad2012330} and CMS~\cite{Chatrchyan:2012aa} experiments. The
ATLAS~\cite{Aad2012142} and CMS~\cite{PhysRevLett.110.022003} experiments also reported
evidence for top-quark associated production with a $W$ boson.  The $s$-channel process
has not yet been observed. Because of the smaller production cross section and larger
backgrounds, it is more difficult to isolate it compared to the $t$-channel process in
proton-antiproton collisions. It is even more difficult at the Large Hadron Collider,
although the absolute production rate is higher, as proton-proton collisions yield a
significantly smaller signal-to-background ratio compared to the Tevatron. Recently the D0
collaboration announced the first evidence for the $s$-channel process in the charge
lepton ($\ell$) +jets channel with a data set corresponding to $9.7~\mathrm{fb}^{-1}$ of
integrated luminosity~\cite{tagkey2013656}.

In this Letter, we present the measurement of the single-top-quark $s$-channel cross
section with the full CDF Run II data set in the $\ell$+jets final state~\cite{thesis}. The data are
collected with the general-purpose Collider Detector at Fermilab (CDF
II)~\cite{PhysRevD.71.032001} and correspond to an integrated luminosity of
$9.4~\mathrm{fb}^{-1}$. The CDF II detector is a solenoid magnetic spectrometer surrounded
by calorimeters and muon detectors.

Since top-bottom quark coupling is much larger than the magnitudes of the top-down and
top-strange quark couplings, we assume that all top quarks decay into \textit{Wb}
pairs. We select events in which the \textit{W} boson decays leptonically into an electron
or a muon with a corresponding neutrino. Electrons or muons from $\tau$ decay are also
accepted. Thus, the final state of the signal process consists of one reconstructed
electron or muon, one corresponding neutrino, and two jets originating from bottom quarks
($b$ jets). Since the final state of this process is the same one as used in the search
for a Higgs boson (\textit{H}) produced in association with a \textit{W}
boson~\cite{PhysRevLett.109.111804}, the techniques used in this Letter are based on this
recent search but with a discriminant optimized for the present measurement.

There are important differences in the jet selection strategy between this
$s$-channel-optimized analysis and the previous measurements~\cite{PhysRevD.82.112005},
which were optimized for the $t$-channel process. The $t$-channel process usually yields
one light-flavor jet in the forward region (pseudorapidity $|\eta|>2.0$), which is crucial to
distinguish the $t$-channel signal from background events. Since including these forward
jets does not lead to a more powerful discriminator for the $s$-channel measurement, only
central jets ($|\eta|<2.0$) are included.  Moreover, for the $s$-channel process, events
with two $b$ jets provide the most sensitivity, while most $t$-channel events have only
one reconstructable $b$ jet. As a result, the sensitivity of the $s$-channel analysis has been
improved with a more efficient $b$-jet selection algorithm~\cite{Freeman201364}.

Events are collected using three classes of online selection requirements (triggers). In
order to improve the lepton acceptance, a novel inclusive trigger strategy is used for
events including a central electron or central muon ($|\eta|<1.0$) with
large transverse momentum \pt~\cite{CDF_coord}.  This improves the trigger efficiency by
$4.7\%$ for electrons and $12.6\%$ for muons compared to the previous single-top-quark
cross section measurement~\cite{PhysRevLett.103.092002}. The details of this technique are
discussed in Ref.~\cite{PhysRevLett.109.111804}. Events triggered by~\MET-based
triggers~\cite{CDF_coord}, which require~\MET~$> 45$ GeV or~\MET~$> 35$ GeV plus two jets,
are also included. These events allow the inclusion of additional identified muon types
and are referred to as the extended muon category~\cite{PhysRevD.82.112005}.

The algorithms for identifying leptons and jets are the same as those used in
Ref.~\cite{PhysRevD.82.112005}. Events passing the trigger requirements are further
selected by requiring exactly one isolated charged lepton with reconstructed transverse
momentum \pt~$> 20$ \gevc. The~\MET~threshold is 20 GeV for events containing central
electrons and extended muons, and 10 GeV for events containing central muons.  Events are
also required to have exactly two jets with transverse energy \et~$> 20$ GeV and
pseudorapidity $|\eta| < 2.0$.  A support-vector-machine
algorithm~\cite{1742-6596-331-3-032045} is used to reduce the contamination from multijet
events that do not contain a $W$ boson.

The invariant mass of the reconstructed top-quark candidate provides the greatest
discrimination between $s$-channel single-top-quark events and non-top-quark
backgrounds. The $z$-component of the neutrino momentum is necessary for the
invariant-mass calculation and can be constrained by implementing the $W$-boson
invariant-mass requirement. We choose the smaller solution when there are two real
solutions to this quadratic equation. As tested with simulated samples, the probability
for this algorithm to yield the correct answer is about 70\%. Correctly selecting the
\textit{b} jet that originated from top-quark decay is necessary to improve the mass
resolution of the reconstructed top quark.  A neural network is employed to select the
correct jet out of the two candidate jets in each event. The neural network uses the
following information on both jets: jet transverse momentum, \pt; invariant mass of the
lepton and one jet, \mlb; invariant mass of the lepton, the neutrino and one jet, \mlvb;
and the jet direction in the off-shell $W$-boson rest frame, $\cos\theta_j$. This
algorithm selects the correct jet in $84\%$ of simulated SM single-top-quark $s$-channel
events.

To further suppress backgrounds, such as light-flavor jets produced in association with a
\textit{W} boson, at least one of the two jets in each event is required to be a $b$ jet.
Because there are several observable properties of $b$ jets that can be used to
discriminate them from other jets, a neural-network tagging algorithm~\cite{Freeman201364}
is used to preferentially select $b$ jets.  Jets are classified based on the output value
of the algorithm as untagged, loose (L), or tight (T) tagged. For T (L) jets, as measured
from simulation, the overall efficiency for selecting true $b$ jets is $42\%\pm1.6$\%
($70\%\pm6.5$\%), while the misidentification rate for charm-quark jets is $8.5\%\pm0.7$\%
($27\%\pm5$\%), and the misidentification rate for jets originating from other quarks and
gluons is $0.89\%\pm0.16$\% ($8.9\%\pm0.9$\%). The tagging efficiency and
misidentification rate applied to each jet depend on the jet \et~and $\eta$. The scale
factors for these variables are also applied to each jet to bring the $b$-tagging
efficiencies in simulation into agreement with those in data. By applying these tagging
requirements to each jet in an event, we construct four non-overlapping tagging
categories: TT, TL, T, and LL. For the double-tag categories, the category with the
highest signal-to-background ratio is chosen if an event satisfies more than one category;
for the single-tag category, one jet of the event is required to be tight tagged and the
other one is untagged.

Signal and background events are modeled using a combination of data-driven methods and
Monte Carlo (MC) simulation including the CDF II detector response modeled by
\textsc{geant3}~\cite{Geant3} with \textsc{cteq5l} parton distribution
function~\cite{CTEQ5L} and tuned to the Tevatron underlying-event
data~\cite{PhysRevD.82.034001}. The single-top-quark events are modeled using
\textsc{powheg}~\cite{POWHEG2009} with the top-quark mass set to 172.5 \gevcc, while quark
shower and hadronization are performed by \textsc{pythia}~\cite{Pythia}. Signal events
generated by \textsc{powheg} are at next-to-leading-order (NLO) accuracy in the strong
coupling $\alpha_s$, which is an improved model compared to the leading-order model used
in Ref.~\cite{PhysRevD.82.112005}. The background model remains unchanged from the
previous measurement~\cite{PhysRevD.82.112005}. The diboson (\textit{WW, WZ, ZZ}),
$t\bar{t}$, and Higgs-boson processes (with the Higgs-boson mass set to 125 \gevcc) are
modeled using simulated events generated with \textsc{pythia} and normalized to the cross
section calculated in Refs.~\cite{PhysRevD.60.113006, ttbar, Higgs}.

Events in which a $W$ or $Z$ boson is produced in association with jets
(\textit{W/Z}+jets) are generated with \textsc{alpgen}~\cite{Alpgen}~\footnote{The
  factorization and renomalization scale in the \textsc{alpgen} samples are both set to be
  $\sqrt{M_{W}^2+\sum\limits_{partons} m_{T}^2}$, where $m_{T}^2=m^2+p_{T}^2/c^2$, $m$ is
  zero for all partons except that $m_b=4.7~\mathrm{GeV/c^2}$, and
  $m_c=1.5~\mathrm{GeV/c^2}$.}, at leading order with up to four partons with
generator-to-reconstructed-jet matching~\cite{1126-6708-2007-01-013,Alwall2008} and their
hadronic shower simulated with \textsc{pythia}.  The background from the multijet process,
which does not contain a $W$ boson, is predicted using a data-driven model.  The
normalizations of multijet and $W$+jets processes are determined in a control sample
(pretag sample) that includes events without any $b$-tag requirement. There are 122 039
events in the pretag sample, which is dominated by $W$+jets and multijet events. Since
multijet events typically have smaller~\MET~than $W$-boson events, their normalizations
are determined by fitting the~\MET~distribution in the control sample. Normalization in
the $b$-tagged signal sample for the $W$+heavy-flavor-jets background is calculated by
applying the tagging efficiency and the fraction of heavy-flavor jets to the rates
calculated in the pretag sample. The fraction of heavy-flavor jets is derived from fitting
jet-flavor-sensitive variables in the $b$-tagged $W$+one-jet data
sample~\cite{PhysRevD.82.112005}. For the $W$+light-flavor background, where one or two
light-quark jets or gluon jets are misidentified as $b$ jets, the normalization is
calculated from the $W$+jets pretag sample by subtracting the heavy-flavor fraction and
multiplying by the per-jet $b$-tag misidentification rate. For the multijet background, a
$b$-tag rate derived from data is used to estimate the normalization of the tagged
multijet background.

\begin{figure}[htbp]
  \includegraphics[width=8.6cm]{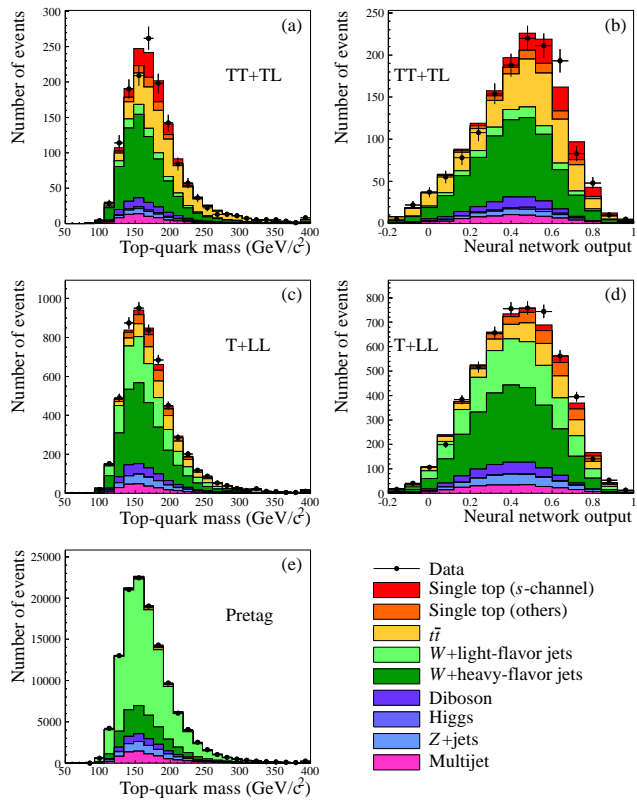}%
  \caption{\label{fig:discr} Distribution of invariant mass of the reconstructed top-quark
candidates (left), and distribution of the neural network output (right). We combine
$b$-tag categories with comparable signal purity: TT+TL for panels (a) and (b), and T+LL
for panels (c) and (d). Events in the pretag control sample are shown in panel (e). Statistical uncertainties are shown for the data points.}
\end{figure}

The estimated event yields are shown in Table~\ref{tab}. Here, and in all following figures,
we combine $b$-tag categories with similar signal purity (TT with TL and T with
LL). Table~\ref{tab} shows that the predicted background and its uncertainty are larger
than the expected signal. By using variables with different distributions for signal and
backgrounds, we improve signal purity in some regions of these distributions. The
invariant mass distribution of top-quark candidates, shown in Fig.~\ref{fig:discr}, is the
most powerful single discriminating variable.

\begin{table}[htbp]
  \begin{center}
    \newcolumntype{d}{D{.}{.}{1}}
    \caption{\label{tab}Summary of background and signal predictions in two summed tagging categories. The predicted uncertainties include statistical and systematic contributions.}
    \begin{tabular*}{8.6cm}{@{\extracolsep{\fill}} l>{\raggedright}d@{\extracolsep{0pt}}@{$\pm$}d@{\extracolsep{\fill}}>{\raggedright}d@{\extracolsep{0pt}}@{$\pm$}d}
      \hline
      \hline
      Category                 & \multicolumn{2}{c}{TT+TL}  & \multicolumn{2}{c}{T+LL} \\
      \hline           
      $t\bar{t}$               & 357    & 40    &  560  & 57   \\  
      Diboson                  & 58.7   & 7.8   &  279  & 34   \\
      Higgs                    & 12.5   & 1.0   &  12.0 & 0.9 \\  
      $Z$+jets                 & 31.6   & 3.5   &  190  & 21   \\  
      Multijet                 & 76     & 31    &  326  & 130  \\  
      $W$+heavy-flavor jet     & 712    & 286   &  2597 & 1046 \\  
      $W$+light-flavor jet     & 66     & 14    &  1220 & 175 \\  
      $t$ and \textit{tW}-channel     & 53.4   & 6.7   &  265  & 30   \\
      \hline 
      $s$-channel              & 116    & 12    &  127  & 12   \\
      \hline 
      Total prediction         & 1484   & 403   &  5574 & 1501\\    
      \hline 
      Observed                 & \multicolumn{2}{c}{1231}       &  \multicolumn{2}{c}{5338}    \\
      \hline
      \hline
    \end{tabular*}        
  \end{center}
\end{table}

We train a set of artificial neural networks~\cite{Hocker:2007ht} to further discriminate
the signal process using the combined information on the reconstructed top-quark mass and
several other variables. The neural networks incorporate the following variables:
invariant mass of the top-quark candidate, $M_{\ell\nu j}$; invariant mass of all signal
final-state particles, $M_{\ell\nu jj}$; transverse momentum of the charged lepton,
$p_T^\ell$; invariant mass of the two jets, $M_{jj}$; angle between the charged-lepton
momentum and the momentum of the jet from the top-quark decay in the top-quark rest frame,
$\cos\theta_{\ell j}$; scalar sum of transverse energy of the two jets, the charged
lepton, and the neutrino, $H_{T}$; transverse mass of the top-quark candidate,
$M_T^{\ell\nu j}$, defined to be the invariant mass calculated using the projections of
the three-momentum components in the plane perpendicular to the beam axis; and output
value of the neural network that determines the $b$ jet most likely to originate from the
top-quark decay. We optimize the neural networks separately for each tagging category and
for different lepton categories using different input variables. The variable
$M_T^{\ell\nu j}$ is used only for extended muon events, and the output value of the
$b$-jet-selector neural network only for the central-lepton events. In the neural network
training, the background samples consist of all backgrounds predicted by simulation, and
the fractional yields among background samples are set as predicted by the background
model.

We use the pretag sample to check the modeling of each input variable. We investigate the
neural-network output in the $b$-tagged signal region only after ensuring that all
variables are well modeled in the control sample. The distributions of neural-network
output are shown in Fig.~\ref{fig:discr}, with categories having similar signal purities
combined.

We employ a binned-likelihood technique to extract the single-top-quark $s$-channel cross
section from the neural-network-output distribution. We assume a uniform prior probability
density for all non-negative values of the cross section and integrate the posterior
probability density over the parameters of effects associated with all sources of
systematic uncertainties, parametrized using Gaussian priors truncated to avoid negative
probabilities.  We include the rate uncertainties from the following sources: $b$-tag
scale factor; charm-quark-jet-misidentification rate; light-flavor-jet-misidentification
rate; luminosity uncertainties; lepton-acceptance uncertainties; theoretical cross section
uncertainties; initial- and final-state radiation; normalization of multijet, $Z$+jets,
and $W$+jets backgrounds; and jet-energy scale. Shape uncertainties on the final
discriminant output that arise from initial-state and final-state radiation, the
jet-energy scale, the renormalization and factorization scales, and the electron multijet
sample are also taken into account. The standard deviation of the expected cross section
distribution obtained from pseudoexperiments is reduced by 17\% if the measurement is
performed without including any of the systematic uncertainties. The most relevant
systematic uncertainties are, in descending order of importance: the luminosity
uncertainties, the $b$-tag scale-factor uncertainties, the normalization of $W$+jets, and
the uncertainties from initial- and final-state radiation.

\begin{figure}[htbp]
\includegraphics[width=8.6cm]{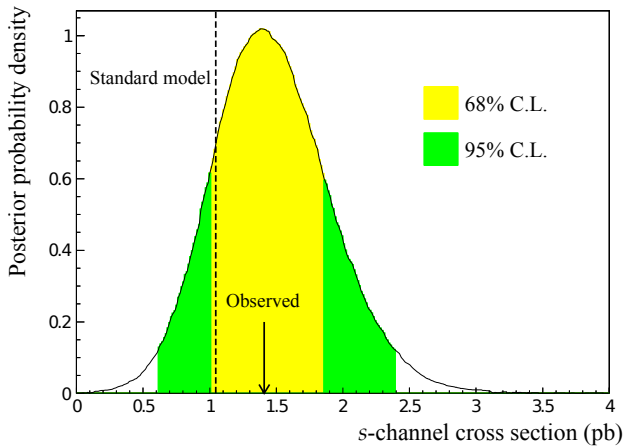}%
\caption{\label{fig:xsec}Posterior probability density distribution for the $s$-channel
cross section measurement, with the SM prediction shown as the vertical dashed line.}
\end{figure} 

The posterior probability density of the $s$-channel cross section is shown in
Fig.~\ref{fig:xsec}.  The most probable $s$-channel cross section is
$1.41^{+0.44}_{-0.42}$~pb, assuming that the top-quark mass is 172.5 \gevcc. This result
is in good agreement with the theory prediction calculated at
next-to-next-to-leading-order accuracy of $1.05 \pm 0.05$~pb~\cite{PhysRevD.81.054028},
which assumes the same top-quark mass, and the uncertainty is assigned to cover the
imperfect estimation of the parton distribution functions. This result also agrees with the
previous measurement from the D0 experiment~\cite{tagkey2013656}. The cross section is
also measured in separate $b$-tagging and lepton categories, and the results in each
independent measurement are all consistent with each other and the theory prediction.

The sensitivity is defined to be the significance expected assuming the SM cross section,
and as measured from pseudoexperiments with the background-only assumption is 2.9 standard
deviations. From background-only pseudoexperiments, we determine the significance of the
excess of the measured cross section over the expected backgrounds as corresponding to a
$p$-value of $5.5\times 10^{-5}$, equivalent to 3.8 standard deviations.  We interpret the
observed excess as evidence of the single-top-quark production through the $s$-channel
process.

In summary, we perform a measurement of the single-top-quark $s$-channel cross section
in the final state with a charged lepton and two jets using the full CDF Run II data
set. We find evidence for the single-top-quark $s$-channel process, and we measure the
$s$-channel cross section to be $1.41^{+0.44}_{-0.42}$~pb, in agreement with the SM
prediction.


We thank the Fermilab staff and the technical staffs of the
participating institutions for their vital contributions. This work
was supported by the U.S. Department of Energy and National Science
Foundation; the Italian Istituto Nazionale di Fisica Nucleare; the
Ministry of Education, Culture, Sports, Science and Technology of
Japan; the Natural Sciences and Engineering Research Council of
Canada; the National Science Council of the Republic of China; the
Swiss National Science Foundation; the A.P. Sloan Foundation; the
Bundesministerium f\"ur Bildung und Forschung, Germany; the Korean
World Class University Program, the National Research Foundation of
Korea; the Science and Technology Facilities Council and the Royal
Society, United Kingdom; the Russian Foundation for Basic Research;
the Ministerio de Ciencia e Innovaci\'{o}n, and Programa
Consolider-Ingenio 2010, Spain; the Slovak R\&D Agency; the Academy
of Finland; the Australian Research Council (ARC); and the EU community
Marie Curie Fellowship Contract No. 302103.


\begin{thebibliography}{32}
\bibitem [{CDF()}]{CDF_coord}%
  \BibitemOpen
  \href@noop {} {}\bibinfo {note} {We use a cylindrical coordinate system with
  the origin at the center of the CDF II detector, $z$ pointing in the
  direction of the proton beam, $\theta$ and $\phi$ representing the polar and
  azimuthal angles, respectively. The transverse energy $E_T$ is defined to be
  $E\sin\theta$. The calorimeter missing $E_T$ ($\METVEC(\mbox{cal})$) is
  defined by the sum over calorimeter towers $\METVEC(\mbox{cal}) = - \sum_{i}
  E_T^i \hat{n}_i$, where $i$ is an index over calorimeter tower with $|\eta| <
  3.6$, and $\hat{n}_i$ is a unit vector perpendicular to the beam axis and
  pointing at the $i$th calorimeter tower. The reconstructed missing energy,
  $\METVEC$, is derived by correcting $\METVEC(\mbox{cal})$ for muon energy
  deposition and jet-energy adjustments. We define $\MET(\mbox{cal})$ and
  $\MET$ to be the scalar magnitudes of $\METVEC(\mbox{cal})$ and $\METVEC$,
  respectively.}\BibitemShut {Stop}%
\end{thebibliography}

%

\end{document}